\documentclass[useAMS]{mn2e}
\usepackage{psfig}
 \usepackage{url}
\usepackage{times}
\def\spose#1{\hbox to 0pt{#1\hss}}
\newcommand\lsim{\mathrel{\spose{\lower 3pt\hbox{$\mathchar"218$}}
     \raise 2.0pt\hbox{$\mathchar"13C$}}}
\newcommand\gsim{\mathrel{\spose{\lower 3pt\hbox{$\mathchar"218$}}
     \raise 2.0pt\hbox{$\mathchar"13E$}}}
\def\sc{Schwarzschild}
\def\ltsima{$\; \buildrel < \over \sim \;$}
\def\lsim{\lower.5ex\hbox{\ltsima}}
\def\gtsima{$\; \buildrel > \over \sim \;$}
\def\gsim{\lower.5ex\hbox{\gtsima}}

\def\sch{Schwarzschild}

\title[SDSS J0131--0321: 11 billion solar mass black hole at z=5.18]
{SDSS J013127.34--032100.1: a candidate blazar with a 11 billion solar mass black hole at $z$=5.18}
\author[G. Ghisellini et al.]
{G. Ghisellini$^1$, \thanks{E--mail: gabriele.ghisellini@brera.inaf.it}
G. Tagliaferri$^1$, T.  Sbarrato$^{2}$, N. Gehrels$^3$  \\
$^1$ INAF -- Osservatorio Astronomico di Brera, via E. Bianchi 46, I--23807 Merate, Italy \\
$^2$ Univ. di Milano Bicocca, Dip. di Fisica G. Occhialini, Piazza della Scienza 3, I--20126 Milano, Italy \\
$^3$ NASA--Goddard Space Flight Center, Greenbelt, Maryland 2077, USA
}

\begin{document}

\pagerange{\pageref{firstpage}--\pageref{lastpage}} \pubyear{2012}

\maketitle
\label{firstpage}

\begin{abstract}
The radio--loud quasar  SDSS J013127.34--032100.1
at a redshift $z$=5.18 is one of the most distant radio--loud objects.
The radio to optical flux ratio (i.e. the radio--loudness) of the source is large,
making it a promising blazar candidate. 
Its overall spectral energy distribution,
completed by the X--ray flux and spectral slope derived through Target of Opportunity 
{\it Swift}/XRT observations, is interpreted by a non--thermal 
jet plus an accretion disc and molecular torus model.
We estimate that its black hole mass is $(1.1\pm0.2)\times 10^{10} M_{\odot}$
for an accretion efficiency $\eta=0.08$, scaling roughly linearly with $\eta$. 
Although there is a factor $\gsim$ 2 of systematic uncertainty,
this black hole mass is the largest found at these redshifts. 
We derive a viewing angle between 3 and 5 degrees.
This implies that there must be other (hundreds) sources with the same black hole 
mass of  SDSS J013127.34--032100.1, but whose jets are pointing away from Earth.
We discuss the problems posed by the existence of such large black hole masses 
at such redshifts, especially in jetted quasars.
In fact, if they are associated to rapidly spinning black holes, the accretion
efficiency is high, implying a slower pace of black hole growth with respect to
radio--quiet quasars.
\end{abstract}

\begin{keywords}
  quasars: general; quasars: supermassive black holes -- X--rays: general
\end{keywords}


\section{Introduction}

Yi et al. (2014, hereafter Yi14) discovered 
SDSS J013127.34--032100.1 (hereafter SDSS 0131--0321) as a radio--loud quasars
a a redshift $z=5.18$ characterised by a large radio
to optical flux ratio (i.e. the so--called radio--loudness) $R\sim 100$.
Yi14 selected the source through optical--IR selection criteria 
based on SDSS and WISE photometric data (Wu et al. 2012), and then observed it
spectroscopically in the optical and the IR.
The IR spectrum 
revealed an absorbed broad Ly$\alpha$ line and a broad MgII line,
that allowed to estimate the black hole
mass of the objects though the virial method, yielding $M_{\rm BH}=(2.7-4)\times 10^9 M_\odot$.
If this were the real black hole mass, then the source would emit above the Eddington limit.

A very large black hole mass found in a jetted quasar
at such large redshifts is somewhat puzzling.
This is because it is commonly believed that the jet is associated
with rapidly spinning black holes, in order to let the Blandford \& Znajek (1977) 
mechanism work.
But if this were the case, the efficiency of accretion would be large,
implying that {\it less} mass is required to emit a given luminosity.
The black hole would then grow at a slower rate, and it becomes problematic
to explain large black hole masses at high redshifts  (see e.g. Ghisellini et al. 2013).
This motivates our interest in jetted quasars at high redshifts
with large black hole masses.

Up to now the three blazars known at $z>5$ are
Q0906+6930 ($z=5.47$; Romani et al. 2004), 
B2 1023+25 ($z=5.3$; Sbarrato et al. 2012, 2013),
and SDSS J114657.79+403708.6 ($z=5.005$; Ghisellini et al. 2014a).
All these sources have a well visible accretion disc emission
in the IR--optical part of the observed spectral energy distribution 
(SED), including strong broad emission lines.
A large  radio--loudness $R$ is a good indicator of the alignment of their jets with the line of sight,
that boosts the non--thermal emission because of relativistic beaming.
Values $R\ge 100$, as in the case of SDSS 0131--0321 (Yi14), 
make the source a good {\it blazar candidate}, i.e. a source whose
jet is seen under a viewing angle $\theta_{\rm v}$ smaller than the beaming angle
$1/\Gamma$, where $\Gamma$ is the bulk Lorentz factor of the jet.
This definition is somewhat arbitrary, but has the merit to divide
in a simple way blazars from their {\it parent population}, i.e. quasars
with jets pointing away from us.
The divide at $\theta_{\rm v} =1/\Gamma$ implies that for each blazar there are other
$2\Gamma^2$ sources of similar intrinsic properties but pointing elsewhere.
However, the radio--loudness alone does not guarantee the classification of the
source as a blazar: $R>100$ could in fact correspond to a source seen at $\theta_{\rm v}>1/\Gamma$,
but with a particularly weak intrinsic optical luminosity, or, vice--versa, with a particular
strong intrinsic radio--luminosity.
To confirm the blazar nature of these high--redshift powerful sources we require
also a strong and hard X--ray flux, because it is an additional signature of a
small viewing angle (due to the specific emission mechanism though to produce the
X--ray emission, see Dermer 1995 and Sbarrato et al. 2015).
For this reason, we asked and obtained a target of opportunity observation
in the X--ray band with the {\it Swift} satellite.
In this letter we discuss these data in the broader context of the 
entire SED, deriving a robust estimate on the black hole mass and 
on the possible range of viewing angles.

In this work, we adopt a flat cosmology with $H_0=70$ km s$^{-1}$ Mpc$^{-1}$ and
$\Omega_{\rm M}=0.3$.

\section{SDSS 0131--0321 as a blazar candidate}

SDSS 0131--0321 was selected by Yi14 as a high--$z$ quasars
on the basis of the Sloan Digital Sky Survey (SDSS) and the
{\it Wide--Field Infrared Survey Explorer (WISE)} photometric data.
The optical and near infrared spectroscopy performed by Yi14 showed 
a broad Ly$\alpha$ and a MgII line, at $z=5.18$.
From the FIRST (Faint Images of the Radio Sky at Twenty-cm; Becker, White \& Helfand, 1995) 
catalog the radio flux is 33 mJy at 1.4 GHz, corresponding to a 
$\nu L_\nu$ luminosity of $1.3\times 10^{44}$ erg s$^{-1}$. 
Yi14 measured a radio loudness of $\sim$100, by assuming that the radio--to--optical spectrum 
follows a $F_\nu\propto \nu^{-0.5}$ power law, and calculating this way the flux
at the rest frame frequencies of 5 GHz and 4400 \AA\ (Kellerman et al. 1989).
The source is not detected by the Large Area Telescope (LAT) onboard the {\it Fermi}
satellite.
In high--$z$ and powerful radio sources, a radio--loudness equal or larger of 100 strongly suggests that
we see the jet radiation with a small viewing angle.
This makes SDSS 0131--0321 a good blazar candidate, but in order to confirm it, we need
to check if its X--ray emission, produced by the jet, is strong relative to the optical, and 
if its slope is harder than the one of a typical accretion disc corona (i.e. $\alpha_X<$0.8--1).

\subsection{{\it Swift} observations}

To confirm the  blazar nature of SDSS 0131--0321 we asked to observe the source with the {\it Swift}
satellite (Gehrels et al. 2004).
In fact the X--ray spectra of blazars beamed towards us are particularly 
bright and hard, with an energy spectral index $\alpha_x \sim 0.5$ [$F(\nu) \propto \nu^{-\alpha_x}$]
(see e.g. Ghisellini et al. 2010, Wu et al. 2013).
The observations were performed between October 23 and December 9, 2014
(ObsIDs:  00033480001--00033480014).

Data of the X--ray Telescope (XRT, Burrows et al. 2005) and the UltraViolet Optical Telescope 
(UVOT, Roming et al. 2005) were downloaded from HEASARC public archive, 
processed with the specific {\it Swift} software included in the package 
{\tt HEASoft v. 6.13} and analysed. 
The calibration database was updated on June, 2014.
We did not consider the data of the Burst Alert Telescope (BAT, Barthelmy et al. 2005), 
given the weak X--ray flux.

From all the XRT observations we extract a single spectrum with total exposure of 20.2 ks.
The mean count rate was  $(2.3 \pm 0.3)\times 10^{-3}$.
%
%
Given the low statistics, the fit with a power law model with Galactic absorption 
($N_{\rm H}=3.7\times 10^{20}$~cm$^{-2}$, Kalberla et al. 2005) 
was done by using the likelihood statistic (Cash 1979).

The output parameters of the model were a spectral slope
$\Gamma_x =(\alpha_x+1)= 1.56 \pm 0.38$ 
and an integrated de--absorbed flux 
$F_{0.3-10\,\rm keV} = (1.4 \pm 0.5) \times 10^{-13}$~erg~cm$^{-2}$~s$^{-1}$. 
The value of the likelihood was 43.9 for 45 dof.
The X--ray data displayed in the SED (Fig. \ref{sed}) have been rebinned 
to have at least $3 \sigma$ in each bin. 

The source was not expected to be detected in any of the UVOT filters, therefore the
various observations were performed with different filters (i.e. the observations were
performed with the set-up ``filter of the day"), usually W1 or M2 UV--filters. 
In four cases
also the $u$ filter was used, while the $b$ and $v$ filters were used once during the observation
of November 12, 2014 for a total exposures of 223 and 512 s, respectively. 
The source was never detected. 
For the $v$ filter we derive a 3$\sigma$ upper limit of $v \sim 20.5$ mag,
corresponding to a flux $F<0.023$ mJy, without accounting for the likely absorption
along the line of sight due to intervening matter.
 
\begin{figure}
\vskip -0.6cm
\hskip -0.3cm
\psfig{file=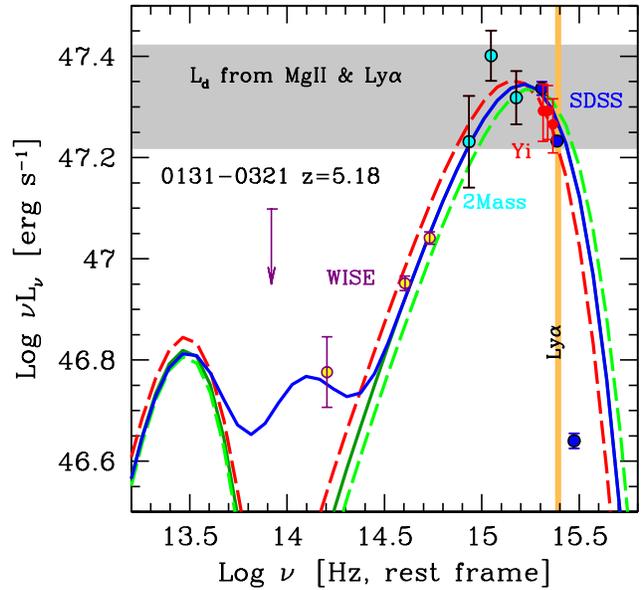,height=9.2cm,width=9.2cm}
\vskip -0.7 cm
\caption{
Optical--UV SED of 0131--0321 in the rest frame, together with 
models of standard accretion disc emission.
The grey stripe indicates the $\nu L_\nu$ peak luminosity of the disc,
estimated as $L_{\rm d}=10\times L_{\rm BLR}$ (see text).
We show the spectrum of three accretion disc models
with different black hole masses:
$M_{\rm BH}/M_\odot=1.4\times 10^{10}$ (red dashed), $1.1\times 10^{10}$ 
(solid blue and dark green) and  $9\times 10^9$ (dashed, light green).
Outside this range of masses, the model cannot fit 
satisfactorily the data. 
The {\it WISE} low frequency point is fitted assuming that, besides the torus emission
envisaged by the Ghisellini \& Tavecchio (2009) model, there is another, hotter ($T\sim$1300 K) 
torus component, as found in Calderone et al. (2013).  
}
\label{zoom}
\end{figure}

\subsection{The accretion disc luminosity}

Fig. \ref{zoom} shows the IR--UV part of the SED,
that can be well described as thermal emission from an accretion disc.
The peak of this thermal emission lies redward of the 
hydrogen Ly$\alpha$ frequency (vertical ocher line), implying that it is 
not an artefact of the absorption due to intervening Ly$\alpha$ clouds.
This allows to derive the luminosity of the disc rather accurately:
$L_{\rm d}=4.1\times10^{47}$ erg s$^{-1}$. 
The (rest frame) peak frequency of the disc emission is $\nu_{\rm d, peak}\sim 1.7\times 10^{15}$ Hz.
Note that $L_{\rm d}$ is roughly half of the bolometric luminosity $L_{\rm bol}$
estimated by Yi14 on the basis of the (rest frame) 3000 \AA\ continuum luminosity, using the
prescription suggested by Richards et al. (2006): $L_{\rm bol}=5.18 L({3000\,{\rm \AA}})$.
As discussed in Calderone et al. (2013), $L_{\rm bol}$ considered
by Richards et al al. (2006) includes the IR ri--emission by the torus
and the X--ray emission, while the ``pure" disc luminosity is $L_{\rm d}\sim 0.5 L_{\rm bol}$.

The {\it WISE} low frequency detection point lies in--between the contribution of the
accretion disc and the torus emission as modelled by Ghisellini \& Tavecchio (2009),
who considered that the torus emits as a simple black--body at the temperature of
$T_{\rm torus} = 370$ K.
In reality, the torus emission is more complex, with emission up to the sublimation dust 
temperature $T_{\rm subl}\sim 2000$ K (e.g. Peterson 1997).
Calderone et al. (2012) found that composite {\it WISE} data of a large sample of quasars
were consistent with the sum of two black--bodies with average temperatures $T$=308 and 1440 K
and similar luminosities.
We have added a black--body with $T=1300$ K to the model SED, having a luminosity
similar to the colder black--body of 370 K.
This scenario is probably still oversimplified, but can account for the
observed data.

Fig. \ref{zoom} shows also that the Two Micron All Sky Survey (2MASS, Skrutskie et al. 2006)
point at $\log (\nu/{\rm Hz})\sim 15.05$ (rest frame)
is above the modelled disc emission.
This discrepancy is due to the MgII broad emission line (at 2800 \AA\, rest frame),
falling in the $H$ band filter of 2MASS (Cohen et al. 2003).

The grey stripe in Fig. \ref{zoom} shows the $L_{\rm d}$ value inferred from the 
Ly$\alpha$ and MgII broad lines.
We assumed that $L_{\rm d}\sim 10 L_{\rm BLR}$ 
(Baldwin \& Netzer, 1978; Smith et al., 1981), where 
$L_{\rm BLR}$ is the luminosity of the entire broad line region, 
that we derive
following the template by Francis et al. (1991):
setting to 100 the contribution of the Ly$\alpha$ line, the broad MgII contribution is 34,
and the entire $L_{\rm BLR}$ is 555.
For the template calculated by Vanden Berk (2001) the contribution of all broad lines
to $L_{\rm BLR}$ is similar, with the exception of the Ly$\alpha$ whose contribution
is about half the one calculated by Francis et al. (1991).

We then took the logarithmic average of the values
derived separately for the MgII and the Ly$\alpha$ line (see data in Yi14),
assuming that we see only half of it (i.e. the non absorbed part).
We assumed an uncertainty of 0.2 dex on this estimate of $L_{\rm d}$.


\subsection{The black hole mass}

We can set a lower limit to the black hole mass of SDSS 0131--0321
by assuming that its disc emits at or below the Eddington level.
This implies $M_{\rm BH}\ge 3.2\times 10^9 M_\odot$.
Yi14, using the virial method (e.g. Wandel 1997; Peterson et al. 2004) 
applied to the MgII broad line, derived $M_{\rm BH}=(2.7-4.0)\times 10^9$,
and concluded that the source is emitting at a super--Eddington rate
(they also considered $L_{\rm bol}$ as the disc luminosity).

\begin{figure}
\vskip -0.6 cm
\hskip -0.3cm
\psfig{file=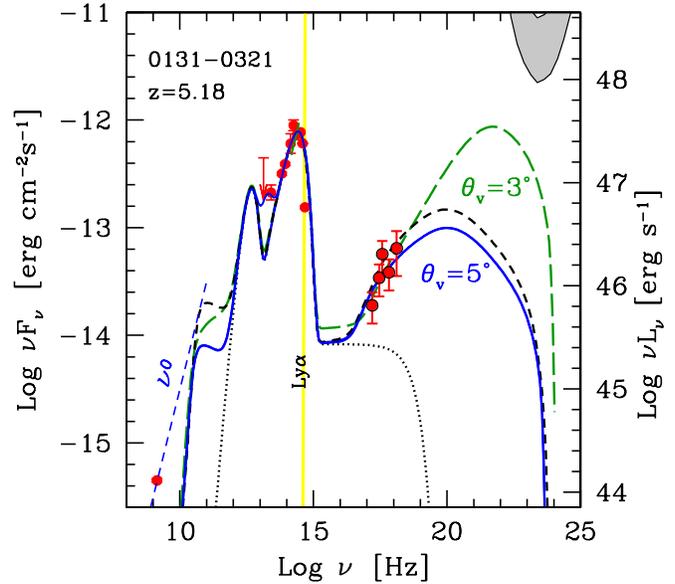,height=9.cm,width=9.cm}
\vskip -0.5 cm
\caption{
SED of SDSS J0131--0321 together with three one--zone leptonic models,
corresponding to $\theta_{\rm v} = 3^\circ$, $\Gamma=13$
(green long dashed line); 
$\theta_{\rm v} = 5^\circ$, $\Gamma=13$
(solid blue line) and
$\theta_{\rm v} = 5^\circ$, $\Gamma=10$
(short dashed black line).
Parameters in Tab. \ref{para}.
The lower bound of the grey stripe correspond to 
the LAT upper limits for 5 years, 5$\sigma$.
} 
\label{sed}
\end{figure}

The IR--optical SED of the source shows a peak and has a low frequency
slope consistent with the emission from a simple, Shakura \& Sunyaev disc (1973) model.
Fig. \ref{sed} shows that the non--thermal, possibly highly variable, 
continuum is not contributing to the IR--UV flux.
Since the accretion disc is much less variable than the jet emission,
we expect that the IR--UV data, although not simultaneous, give a good description
of the disc emission.
This depends only on $M_{\rm BH}$ and the mass accretion rate $\dot M$.
The latter is traced by the total disc luminosity $L_{\rm d}=\eta\dot{M}c^2$,
where $\eta$ is the efficiency, that depends on the last stable orbit, hence
on the black hole spin  (with $\eta=0.057$ 
and $\eta=0.3$ appropriate for a non rotating 
and a maximally spinning black hole, respectively; Thorne 1974).

Having already measured $L_{\rm d}$ (hence $\dot M$, assuming $\eta\sim 0.08$), 
the only free parameter is $M_{\rm BH}$, whose value determines the peak frequency 
of the disc emission.
For a fixed $\dot M$, a larger $M_{\rm BH}$ implies a larger disc surface, hence a lower 
maximum temperature. 
Therefore a larger $M_{\rm BH}$ shifts the peak to lower frequencies. 
Then the best agreement with the data fixes $M_{\rm BH}$ (e.g. Calderone et al. 2013).
Fig. \ref{zoom} shows the disc emission for three (slightly) different 
values of $M_{\rm BH}$: 9, 11 and 14 billions solar masses.
Outside this range we do not satisfactorily account for the
observed data, while the agreement between the data and the model is
rather good within this mass range.
We conclude that the black hole mass is well determined
and is $M_{\rm BH}=(1.1\pm 0.2)\times 10^{10} M_\odot$
for $\eta=0.08$.
We obtain $M_{\rm BH}=8\times 10^{9} M_\odot$
for $\eta=0.057$ and $M_{\rm BH}=2\times 10^{10} M_\odot$ for 
$\eta=0.15$ (see also \S 4).
There is thus a factor of $\gsim 2$ of systematic uncertainty.

\begin{table*} 
\centering
\begin{tabular}{llllllllllllllllll}
\hline
\hline
Name   &$z$ &$R_{\rm diss}$ &$M_{\rm BH}$ &$R_{\rm BLR}$ &$P^\prime_{\rm i}$ &$L_{\rm d}$ &${L_{\rm d}\over L_{\rm Edd}}$ 
&$B$ &$\Gamma$ &$\theta_{\rm v}$
    &$\gamma_{\rm b}$ &$\gamma_{\rm max}$  &$P_{\rm r}$ &$P_{\rm B}$ &$P_{\rm e}$ &$P_{\rm p}$ \\
~[1]      &[2] &[3] &[4] &[5] &[6] &[7] &[8] &[9] &[10] &[11]  &[12] &[13]  &[14] &[15] &[16] &[17] \\
\hline   
0131--0321 &5.18  &2310  &1.1e10 &2031 &0.02  &47.62 &0.25  &1.6 &13  &5  &70  &3e3   &46.3 &47.2 &44.9 &47.6 \\
0131--0321 &5.18  &2310  &1.1e10 &2031 &0.02  &47.62 &0.25  &2.1 &10  &5  &70  &3e3   &46.0 &47.2 &44.8 &47.4 \\
0131--0321 &5.18  &2310  &1.1e10 &2031 &9e--3 &47.62 &0.25  &1.1 &13  &3  &300 &3e3   &46.0 &46.9 &44.0 &46.5 \\
\hline
1146+430   &5.005 &900  &5e9    &1006 &7e--3 &47.00 &0.15  &1.4 &13  &3  &230 &3e3   &45.9 &46.3 &44.0 &46.5 \\ 
0906+693   &5.47  &630  &3e9    &822  &0.02  &46.83 &0.17  &1.8 &13  &3  &100 &3e3   &46.3 &46.2 &44.6 &47.1  \\   
1023+25    &5.3   &504  &2.8e9  &920  &0.01  &46.95 &0.25  &2.3 &13  &3  &70  &4e3   &46.0 &46.2 &44.5 &46.9   \\    
\hline
\hline 
\end{tabular}
\caption{List of parameters adopted for or derived from the model for the SED of 0131--0321, 
compared with the set of parameters used for the other two blazars at $z>5$.
Col. [1]: name;
Col. [2]: redshift;
Col. [3]: dissipation radius in units of $10^{15}$ cm;
Col. [4]: black hole mass in solar masses;
Col. [5]: size of the BLR in units of $10^{15}$ cm;
Col. [6]: power injected in the blob calculated in the comoving frame, in units of $10^{45}$ erg s$^{-1}$; 
Col. [7]: logarithm of the accretion disk luminosity (in erg s$^{-1}$);
Col. [8]: $L_{\rm d}$ in units of $L_{\rm Edd}$;
Col. [9]: magnetic field in Gauss;
Col. [10]: bulk Lorentz factor at $R_{\rm diss}$;
Col. [11]: viewing angle in degrees;
Col. [12] and [13]: break and maximum random Lorentz factors of the injected electrons;
Col. [14]--[17]: logarithm of the jet power (in erg s$^{-1}$) in different forms: Col. [14]
  power spent by the jet to produce the non--thermal beamed radiation; Col. [15]: jet Poynting flux;
  Col. [16]: power in bulk motion of emitting electrons; Col. [17]: power in bulk motion of cold protons,  
  assuming one proton per emitting electron.
The total X--ray corona luminosity is assumed to be in the range 10--30 per cent of $L_{\rm d}$.
Its spectral shape is assumed to be always $\propto \nu^{-1} \exp(-h\nu/150~{\rm keV})$.
}
\label{para}
\end{table*}

\section{Overall spectral energy distribution}

The overall SED of SDSS 0131--0321 from radio to X--rays is shown in Fig. \ref{sed}.
To guide the eye, the dashed line in the radio domain show a radio spectrum $F_\nu \propto \nu^0$.
The 5 years limiting sensitivity of {\it Fermi}/LAT is shown by the lower boundary of the grey hatched area.
The solid and long--dashed lines correspond to the one--zone, leptonic model
described in detail in Ghisellini \& Tavecchio (2009).
The model assumes most of the emission is produced at a distance $R_{\rm diss}$ from the black hole,
where photons produced by the BLR and by the torus  are the most important seeds for the
inverse Compton scattering process, that dominates the high energy luminosity.
The region moves with a bulk Lorentz factor $\Gamma$ and has a tangled magnetic field $B$.
The distribution of the emitting electrons is derived through a continuity equation,
assuming continuos injection, radiative cooling, possible pair production
and pair emission, and is calculated at a time $r/c$ after the start of the injection, 
where $r=\psi R_{\rm diss}$ is the size of the source, and $\psi$ (=0.1 rad) is the
semi--aperture angle of the jet, assumed conical.
The accretion disk component is accounted for, as well the
infrared emission reprocessed by a dusty torus and the X--ray 
emission produced by a hot thermal corona sandwiching the accretion disc.

The three models shown in Fig. \ref{sed}, whose parameters are listed in Tab. 
\ref{para}, differ mainly from the value of $\theta_{\rm v}$ and $\Gamma$.
This highlights how the predicted SED changes, especially in the hard X--ray band, by
varying $\theta_{\rm v}$ even by a small amount.
The two $\theta_{\rm v}=5^\circ$ cases 
indicate the maximum viewing angle that can well explain
the existing data.
The $\theta_{\rm v}=5^\circ$, $\Gamma=10$ and the $\theta_{\rm v}=3^\circ$ cases 
correspond to a viewing angle smaller
than $1/\Gamma$, that would allow to classify SDSS 0131--0321 a blazar.
To refine our possible choices
we need observations in the hard X--ray band, such as the one
that the {\it Nuclear Spectroscopic Telescope Array} satellite ({\it NuSTAR}, Harrison et al.\ 2013) can provide.
As in the case of B2 1023+25 (Sbarrato et al. 2013), {\it NuSTAR} could reveal a
very hard spectrum, that could make us to choose the large $\Gamma$ and small $\theta_{\rm v}$ solution.
Tab. \ref{para} reports also,
for the ease of the reader, the parameters we used to fit the other 3 known 
blazars\footnote{For the jet powers, we here consider the sum of power of each of the two jets,
while in the original papers discussing these sources, the power of only one jet was reported.} at $z>5$.
The parameters are all very similar, and are also in the bulk of the distribution of parameters
accounting for much larger sample of blazars showing broad emission lines
(Ghisellini et al. 2010; 2014b, Ghisellini \& Tavecchio 2015).

\begin{figure}
\vskip -0.6 cm
\hskip -0.3cm
\psfig{file=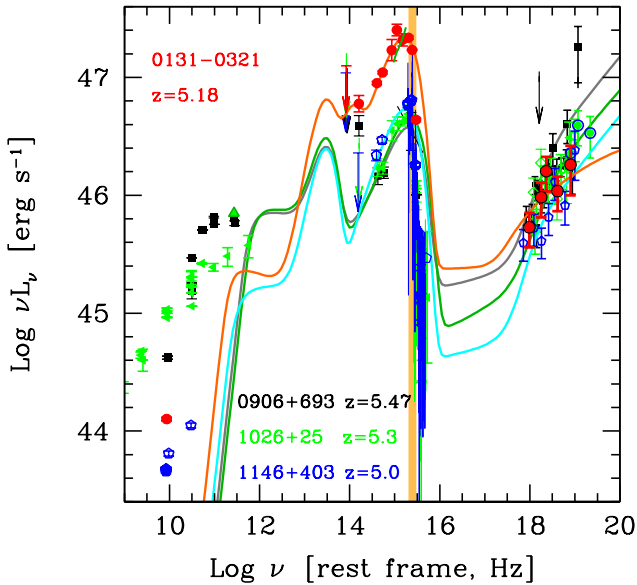,height=9cm,width=9cm}
\vskip -0.5 cm
\caption{The SED of SDSS 0131--0321 is compared to the SEDs of the other
known blazars at $z\ge$5.
SDSS 0131--0321 stands out by having the most powerful disc.
The radio and the X--ray data, on the other hand, have  
luminosity similar to the other blazars.
The figure shows also the interpolating models, whose parameters
are listed in Tab. \ref{para}.
} 
\vskip -0.2 cm
\label{comparison}
\end{figure}

Fig. \ref{comparison} shows how the SED of SDSS 0131--0321 compares with the SED
(in $\nu L_\nu$ vs rest frame $\nu$) of the three other blazar known at $z>5$.
The 4 non--thermal jet SEDs are rather  similar,
but the accretion disc component of SDSS 0131--0321 stands out, being
a factor $\sim$4 more luminous.

\section{Discussion and conclusions}

Although we cannot decide (yet) if SDSS 0131--0321 can be classified
as a blazar in a strict sense (i.e. a source with $\theta_{\rm v}<1/\Gamma$),
we can nevertheless conclude that $\theta_{\rm v}$ is small,
and therefore it is very likely that there exist other sources 
like SDSS 0131--0321 pointing in other directions.
Assuming two oppositely directed jets, $\Gamma=13$ and $\theta_{\rm v}=5^\circ$,
we can estimate the presence of other $1/(1-\cos 5^\circ)=260$ sources
sharing the same intrinsic properties of SDSS 0131--0321 in the same
sky area (about 1/4 of the entire sky) covered by the SDSS+FIRST survey.
If future {\it NuSTAR} observations detect a high hard X--ray flux, 
then the $\Gamma=13$, $\theta_{\rm v}=3^\circ$ solution is preferred, and the
total number of sources like SDSS 0131--0321 should be  
$2\Gamma^2=338 (\Gamma/13)^2$ in the same  sky area.

We believe that the black hole mass we found through the disc--fitting method is
very reliable, since we do see the peak of the accretion disc emission red--ward of
the Ly$\alpha$ limit.
The only uncertainty concerns the use of a simple Shakura and Sunyaev (1973) 
model, that assumes a no (or a modestly) spinning black hole.
On the other hand the resulting fit is good, and we stress that in the case of a fastly 
spinning black hole, with a corresponding high accretion efficiency $\eta$,
we would derive a {\it larger} black hole mass. 
This is because, for a given $M_{\rm BH}$ and $\dot M$, a Kerr black hole would produce more optical--UV flux
(with respect to a non spinning black hole),
then exceeding the observed data points at the peak of the optical--UV SED.
One must lower $\dot M$, but this leads to under--represent
the IR--optical point, that can be recovered by increasing the mass (i.e. increasing the 
disc surface, implying a smaller temperature).

\begin{figure}
\vskip -0.6 cm
\hskip -0.3cm
\psfig{file=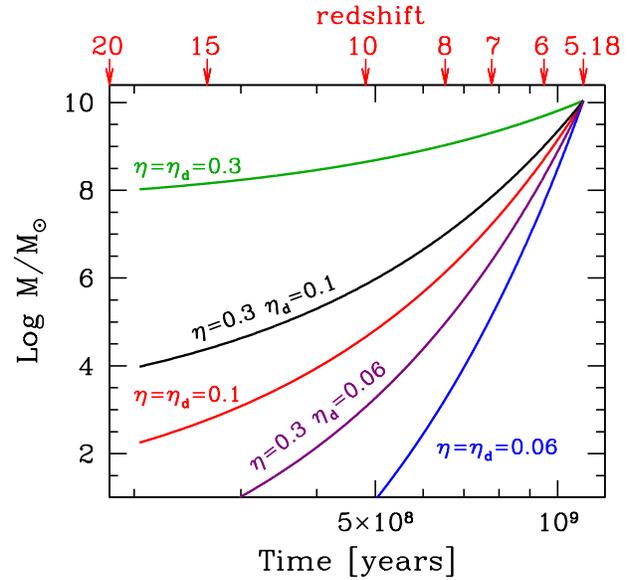,height=9cm,width=9cm}
\vskip -0.5 cm
\caption{The black hole mass as a function of time in system characterised by
an Eddington limited accretion rate, and by different values 
of the efficiency $\eta$ (see also Fig. 3 in Trakhtenbrot et al. 2011) 
Values of $\eta=0.06$ and $\eta=0.3$ correspond to no spinning \sch\ and maximally
spinning Kerr black holes, respectively.
All cases corresponds to $M_{\rm BH} =1.1\times 10^{10} M_\odot$ at $z=5.18$, the
mass of SDSS 0131--0321. 
This figure shows that only if the {\it accretion disc efficiency $\eta_{\rm d}$
is small} we can grow a large mass black hole in the required time starting from
a reasonable seed.  
On the other hand, this may not imply a non rotating black hole.
} 
\label{salpeter}
\end{figure}

How can such a large mass be produced at $z=5.18$?
At this redshift the Universe is 1.1 Gyr old.
Fig. \ref{salpeter} shows the change of the black hole mass
in time, assuming different efficiencies $\eta$.
If the black hole is not spinning, and $\eta<0.1$, then it is possible to grow a black hole 
up to 11 billion solar masses starting from a $100 M_\odot$ seed if the accretion proceeds 
at the Eddington rate all the time.
But if $\eta=0.3$, appropriate for a maximally spinning and accreting black hole (Thorne 1974),
then the growth is slower, and an Eddington limited accretion cannot produce a $1.1\times 10^{10} M_\odot$
black hole at $z=5$, unless the seed is $10^8 M_\odot$ at $z=20$.

This poses the problem: jetted sources are believed to be associated with
fastly spinning black holes, therefore with highly efficient accretors.
If the accretion is Eddington limited, jetted sources should
have black holes {\it lighter} than radio--quiet quasars with not--spinning black holes.
The solution to this puzzle can be that, when a jet is present, then not all
the gravitational energy of the infalling matter is transformed into heat and then 
radiation, as suggested by Jolley \& Kunzic (2008) and 
Ghisellini et al. (2010). 
In this case the {\it total} accretion efficiency can be $\eta=0.3$, but only a fraction of it
($\eta_{\rm d}$) goes to heat the disc, while the rest ($\eta-\eta_{\rm d}$) goes to 
amplify the magnetic field necessary to launch the jet (by tapping the rotational energy of the black hole). 
The disc luminosity becomes Eddington limited for a greater accretion rate (making the black hole growing faster).
This is shown in Fig. \ref{salpeter} as the case $\eta=0.3$, $\eta_{\rm d}=0.1$,
that requires a seed of $10^4 M_\odot$ at $z=20$; and by the case
$\eta=0.3$, $\eta_{\rm d}=0.06$, that can reach 11 billion $M_\odot$ starting from a
$100 M_\odot$ seed at $z\sim$11.

In this scenario, the jet {\it is required} if very large masses must be reached
at large redshifts, together with a vast reservoir of mass that can be accreted.
In this respect, the recent finding of Punsly (2014), namely that radio--loud objects
have a {\it deficit} of UV emission with respect to radio--quiet quasar of similar
disc luminosity, is very intriguing, pointing to the possibility that the inner
part of the accretion disc around a Kerr black hole is producing much less
luminosity than expected, because the produced energy does not go into heating the disc, but
in other forms, such as amplifying the magnetic field of the inner disc,
a necessary ingredient to let the Blandford \& Znajek process work efficiently.

 
\section*{Acknowledgements}
This publication makes use of data products from the Wide--field Infrared Survey Explorer, 
which is a joint project of the University of California, Los Angeles, and the Jet Propulsion
Laboratory/Caltech, funded by NASA.
It also makes use of data products from the Two Micron All Sky Survey, which is a joint project 
of the University of Massachusetts and the Infrared Processing and Analysis Center/Caltech, 
funded by NASA and the National Science Foundation. 
Part of this work is based on archival data and on--line service provided by the ASI 
Science Data Center (ASDC).

\label{lastpage}
\end{document}